\documentclass[a4paper,onecolumn,12pt,assc]{IEEEtranASSC}

\usepackage{graphicx}
\usepackage{amsmath}
\usepackage{url}
\usepackage{bm}
\usepackage{caption}
\usepackage{subcaption}
\usepackage{epstopdf}
\ifCLASSINFOpdf
\else
\fi
\begin{document}
%
\title{Position and Velocity estimation of Re-entry Vehicles using Fast Unscented Kalman Filters}


\author{\IEEEauthorblockN{Sanat Biswas\IEEEauthorrefmark{1},
Li Qiao\IEEEauthorrefmark{2},
Andrew Dempster\IEEEauthorrefmark{1}}

\IEEEauthorblockA{\IEEEauthorrefmark{1}
Australian Centre for Space Engineering Research,
UNSW Australia, NSW, Australia, 2052}

\IEEEauthorblockA{\IEEEauthorrefmark{2}
School of Engineering and Information Technology,
UNSW Australia, Canberra, ACT, Australia, 2600}}

\maketitle

\begin{abstract}
Accurate position and velocity estimation of a re-entry vehicle is essential for realizing its deviation from the desired descent trajectory and providing necessary guidance command in real-time. Generally the Extended Kalman Filter (EKF) is utilized for position and velocity estimation of a space vehicle. However, in the EKF the error covariance is predicted by linearizing the non-linear dynamic model of the system, which results in less accurate state estimation when the dynamics is highly non-linear. As the dynamics of a re-entry vehicle is particularly non-linear in nature, a more accurate position and velocity estimation is expected using a non-linear estimator. The Unscented Kalman Filter (UKF) predicts the mean state vector and the error covariance by deterministic sampling and utilizing the non-linear dynamics of the system. This results in better estimation accuracy than the EKF. However, the processing time of the UKF is much higher than the EKF because of the requirement of multiple state propagations in each measurement time interval. In this paper, application of two new UKF based estimation techniques with reduced processing time in re-entry vehicle position and velocity estimation problem using ground-based range and elevation measurements is presented. The first method is called the Single Propagation Unscented Kalman Filter (SPUKF) where, the \textit{a postiriori} state is propagated only once and then the sampled sigma points at the next time state are approximated by the first-order Taylor Series terms. In the second method called the Extrapolated Single Propagation Unscented Kalman Filter (ESPUKF), the sigma points are approximated to the second-order Taylor Series terms using the Richardson Extrapolation.  The EKF, SPUKF, ESPUKF and the UKF are utilized in a re-entry vehicle navigation scenario using range and elevation measurements. The estimation accuracies and the processing times for different algorithms are compared for the scenario. The result demonstrates that the UKF provides better accuracy than the EKF but requires more processing time. The SPUKF accuracy is better than the EKF and the processing time is significantly less than the UKF. However, the accuracy of the SPUKF is less than the UKF. The ESPUKF provides estimation accuracy comparable to the UKF and the processing time is also significantly reduced.    
\end{abstract}


\section*{Introduction}
Estimation of the position and velocity of a re-entry vehicle is a challenging task due to the highly non-linear nature of the vehicle dynamics. Accurate position and velocity estimation is essential for proper re-entry procedures and vehicle recovery \cite{lu1997entry, johnson2001reusable}. Generally for a re-entry mission, the position and velocity information of the vehicle is estimated from the radar based observations using Kalman Filter.

Out of several types of sequential estimators, the Kalman Filter is designed for the state estimation of stochastic dynamic systems \cite{Bar-shalom2004}. It is the most popular estimation technique due to the computational efficiency and frequently used in space vehicle navigation and attitude estimation. 

The Kalman Filter is a statistical approach to optimal state estimation for linear systems and measurements with random noise \cite{Kalman1960}. The Extended Kalman Filter (EKF) was developed to apply the Kalman Filter framework in non-linear systems \cite{Smith1962, Smith1964}. Application of the EKF spans almost all the engineering disciplines. However, this algorithm provides sub-optimal estimation for mildly non-linear problems \cite{Cox1964, Athans1968} due to the first-order Taylor series approximation of the mean and conditional error covariance \cite{Bernstein1966}. It is long established that the degree of non-linearity of a dynamic system is one of the decisive factors for the accuracy of the EKF. To address the non-linearity, several techniques involving analytical and numerical computation of the Jacobian and Hessian were developed \cite{Athans1968, NoRgaard2000}. Julier et al. suggested a deterministic sampling approach to compute the \textit{a priori} mean state vector and the error covariance to capture the non-linearity of the dynamic system \cite{Julier2000,Julier1997,Julier1998,Julier2003, Julier2004}. This approach is known as the Unscented Kalman Filter.
The estimation accuracy is significantly better than the EKF for systems with Non-linearity Index higher than 0.7 \cite{Biswas2016c}.

The UKF relies on propagation of multiple sample state vectors to predict the \textit{a priori} mean state vector and the error covariance at an epoch. This requires a substantial amount of processing time compared to the EKF. To reduce the computation time of the UKF, two new UKF based estimation techniques called the Single Propagation Unscented Kalman Filter (SPUKF) and the Extrapolated Single Propagation Unscented Kalman Filter (ESPUKF) were proposed in \cite{Biswas2016}. In these new methods only one sample state vector is propagated and the other samples are computed using the Taylor Series approximation. These new filters were applied to a vertical re-entry problem and a LEO satellite navigation problem and the performance were compared with the EKF and the UKF. In \cite{Biswas2016b} the EKF, UKF, SPUKF and ESPUKF was applied to a launch vehicle trajectory estimation problem using GNSS observations. The ESPUKF provided the most optimal estimation performance in terms of the processing time and the estimation accuracy.

In this paper, the SPUKF and the ESPUKF are applied to a realistic re-entry mission. A detailed re-entry vehicle motion considering both the horizontal and vertical motion is simulated. The radar range and elevation observations corresponding to the trajectory is generated. These observations are used in the EKF, UKF, SPUKF and the ESPUKF for the position and the velocity estimation. The performance of the SPUKF and the ESPUKF are compared with the EKF and the UKF. 
\section*{Re-entry Vehicle Dynamics}
\begin{figure}%
\centering
\includegraphics[width=0.7\columnwidth]{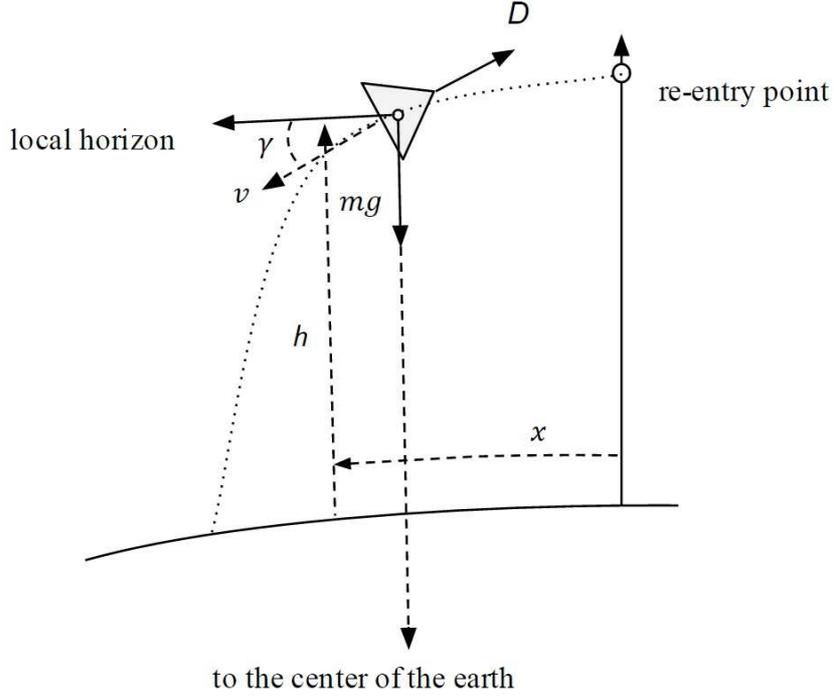}%
\caption{Re-entry vehicle trajectory}%
\label{fig:RV_diag}%
\end{figure}

The re-entry vehicle follows a curved path during atmospheric re-entry. In a detailed re-entry dynamics the downrange $x$, altitude $h$, velocity $v$, flightpath angle $\gamma$ and the aerodynamic co-efficient $C$ are considered as state variables. A planer re-entry vehicle trajectory is provided in Fig. \ref{fig:RV_diag}. 

The re-entry dynamics can be expressed as \cite{regan1993dynamics}:
\begin{equation}
\begin{bmatrix}
\dot x\\
\dot h\\
\dot v\\
\dot \gamma\\
\dot C
\end{bmatrix} = \left[\begin{array}{l}
								\frac{R_E}{R_E + h}v\cos{\gamma}\\
								v\sin{\gamma}\\
								- \frac{D}{m} - g\sin{\gamma}\\
								-\frac{1}{v}\left(g - \frac{v^2}{R_E + h}\right)\cos{\gamma}\\
								0
								\end{array}\right] + \bm{\nu}(t)
\label{eq:d_rentry}
\end{equation} 
where $m$ is the mass of the re-entry vehicle, $R_E$ is the mean radius of the Earth, $g$ is the gravitational acceleration, $D$ is the aerodynamic drag and $\bm \nu(t)$ is the process noise vector. Similar to the launch vehicle model, $D$ is modelled using the following equation \cite{Curtis2010}:

\begin{equation}
D = \frac{1}{2}AC\rho_0e^{-\frac{h}{H}}v^2
\label{eq:drag_f}
\end{equation}

\subsection*{Reference Trajectory Generation}
\begin{figure}[h]%
\centering
\begin{subfigure}{0.48\textwidth}
\includegraphics[width=\textwidth]{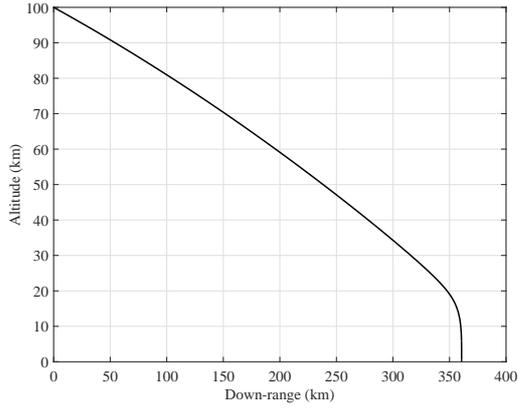}%
\caption{Trajectory}%
\label{fig:crv_traj}%
\end{subfigure}
~
\begin{subfigure}{0.48\textwidth}%
\includegraphics[width=\textwidth]{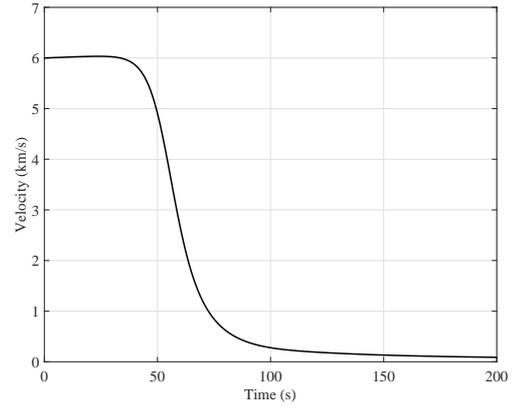}%
\caption{Velocity profile}%
\label{fig:crv_vel}%
\end{subfigure}
\caption{Re-entry vehicle trajectory and velocity profile}
\end{figure}

\begin{figure}[h]%
\centering
\begin{subfigure}{0.48\textwidth}
\includegraphics[width=\textwidth]{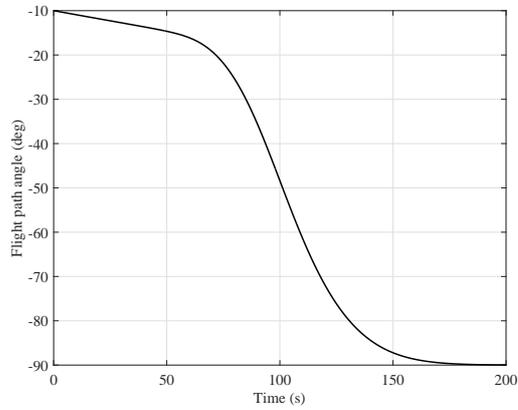}%
\caption{Flightpath angle}%
\label{fig:crv_fpa}%
\end{subfigure}
~
\begin{subfigure}{0.48\textwidth}%
\includegraphics[width=\textwidth]{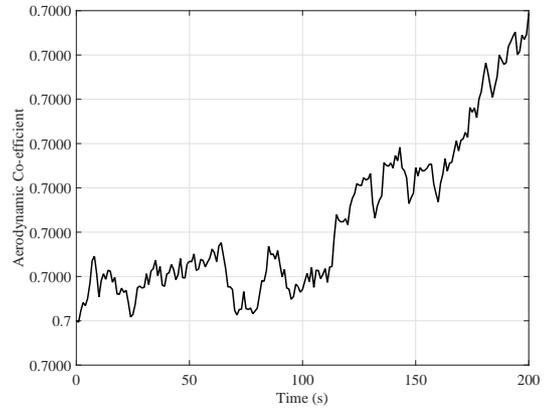}%
\caption{Aerodynamic co-efficient}%
\label{fig:crv_C}%
\end{subfigure}
\caption{Flightpath angle and aerodynamic co-efficient profile}
\end{figure}

The reference re-entry trajectory is generated by numerically integrating equation (\ref{eq:d_rentry}). The initial true state vector is
\begin{equation}
\begin{bmatrix}
h \\ d \\ v \\ \gamma \\ C
\end{bmatrix} = \begin{bmatrix}
								100~km\\ 0~km \\ 6~km/s \\ -10^{\circ}\\ 0.7 \end{bmatrix}
\label{eq:rv_ini}
\end{equation}
The reference trajectory, velocity, flightpath angle and aerodynamic co-efficient profile are shown in figures \ref{fig:crv_traj}, \ref{fig:crv_vel}, \ref{fig:crv_fpa} and \ref{fig:crv_C}.

\section*{Simulation of Radar Observations}
Radar observations are used in the launch vehicle and re-entry vehicle state estimation scenarios. It is assumed that the radar is situated in the trajectory plane of the vehicle and hence the azimuth angle is fixed.

\begin{figure}[h]%
\centering
\includegraphics[width=0.6\columnwidth]{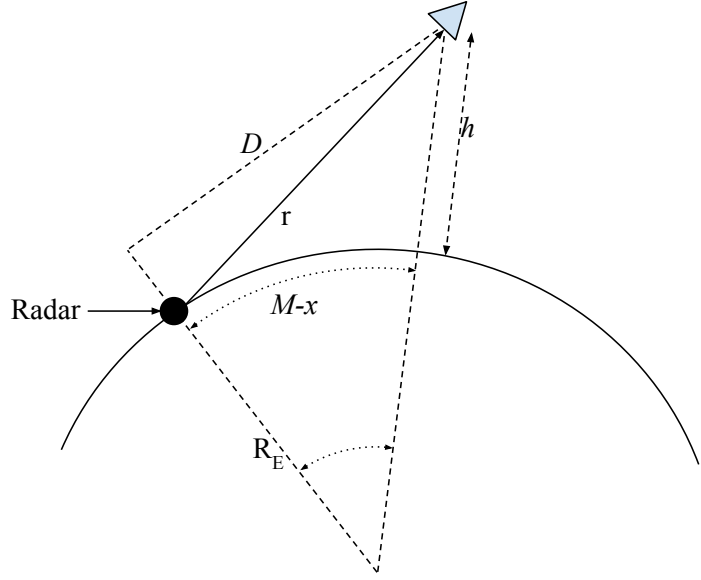}%
\caption{Radar tracking geometry for a re-entry vehicle in a curved trajectory}%
\label{fig:crv_obs}%
\end{figure}

For the curved re-entry trajectory the range and elevation are considered as the observations. The effect of the Earth's curvature is considered in the measurement simulation. The geometry of radar tracking for a re-entry vehicle in a curved path is shown in Fig. \ref{fig:crv_obs}.
 
In the figure, $r$ is the geometric distance of the re-entry vehicle, $M$ is the initial down-range distance of the radar from the re-entry point, $h$ is the altitude and $x$ is the downrange distance of the re-entry vehicle from the re-entry point. $\phi$ is the angular distance between the radar and the re-entry vehicle and can be written as
\begin{equation}
\phi = \frac{M-x}{R_E}
\label{eq:crv_phi}
\end{equation} 
$\delta H$ and $\delta D$ can be expressed as
\begin{align}
\delta H(t) &= (R_E+h(t))\sin\phi -R_E\label{eq:h_diff}\\
\delta D(t) &= (R_E+h(t))\sin\phi\label{eq:d_diff}
\end{align}
The range $r$ and elevation $E$ is modeled as
\begin{align}
r(t) &= \sqrt{\delta H(t)^2 + \delta D(t)^2} + \omega_r(t) \label{eq:crv_range}\\
E(t) &= \tan^{-1}\frac{\delta D(t)}{\delta H(t)} + \omega_E(t) \label{eq:crv_elevation}
\end{align}
here, $\omega_r(t)$ and $\omega_E(t)$ are zero mean white noise with standard deviation of 20m and 17.5 milirad respectively. The range and elevation measurements are generated using equations (\ref{eq:crv_range}) and (\ref{eq:crv_elevation}).
\section*{Implementation of Unscented Filters}
In unscented filtering, the evolution of the process noise statistics over time is addressed by augmenting the state vector with the process noise terms \cite{Julier1997}. The augmented state vector is
\begin{equation}
\bm{X}_a(t) = \begin{bmatrix}
											  \bm{X}(t)\\
												\bm{\nu}(t)
											\end{bmatrix}	
\label{eq:augmented}
\end{equation}
In the UKF, the sigma points are calculated from \cite{Julier2000}
\begin{equation}
\bm{X}^+_a(t) = \begin{bmatrix}
													\bm{X}^+(t)\\
													\bm{0}_{8\times 1}
													\end{bmatrix}
\label{eq:augmented_0}
\end{equation}
\begin{equation}
\bm{P}_a(t) = \left[\begin{array}{ccc}
\bm{P}(t) & \bm{P}_{X\nu}(t)\\
\bm{P}_{X\nu}(t) & \bm{Q}(t)
\end{array}\right]
\label{eq:augmented_cov}
\end{equation} 
Here, $\bm{X}^+(t)$ and $\bm{X}^+_a(t)$ are the \textit{a posteriori} state vector and the augmented state vector respectively at epoch $t$. The augmentation terms are zero because the process noise distribution is considered as zero mean Gaussian. $\bm{P}(t)$ and $\bm{P}_a(t)$ are the error covariance and the augmented error covariance matrices respectively. $\bm{P}_{X\nu}(t)$ is the cross covariance of $\bm X$ and $\bm \nu$. $\bm Q(t) = \bm E[\bm \nu \bm \nu^T]$ is the process noise covariance matrix. The dimension of the augmented state vector is 10. Therefore, a total of 21 sigma points must be propagated to the next epoch to predict the weighted \textit{a priori} mean state vector and the error covariance. The sigma points and the corresponding weights are
\begin{align}
\bm X_0(t) &= \bm{X}^+_a(t)\\
\bm X_i(t) &= \bm{X}^+_a(t) + \bm{\Delta X}_i, (i = 1,2,3...32)\\
W_0 &= \frac{\kappa}{n+\kappa}\\
W_i &= \frac{1}{2(n+\kappa)}, (i = 1,2,3...32)\label{eq:sigma_point}
\end{align}
and
\begin{center}
\begin{tabular}{lcl}
$\Delta{\bm{X}}_i$ &= $(\sqrt{(n+\kappa){\bm{P}}}_a)_i$ & for $i = 1,2,3....16$\\
$\Delta{\bm{X}}_i$ &= $-(\sqrt{(n+\kappa){\bm{P}}}_a)_i$ & for $i = 17,2,3....32$\\
\end{tabular}
\end{center}
where $(\sqrt{(n+\kappa)\bm{P}_a})_i$ is the $i$th column of the matrix $\sqrt{(n+\kappa)\bm{P}_a}$.
$\kappa$ is a parameter and generally it is selected in such a way that $(n+\kappa) = 3$\cite{Julier2000}. Corresponding to all the 21 propagated sigma points the measurement vectors are computed using the measurement equations (\ref{eq:crv_range}) and (\ref{eq:crv_elevation}). The weighted mean of these is considered to be the predicted measurement vector. The measurement error covariance and the cross covariance between the measurement vector and the state vector is computed using the predicted mean state and measurement vector, the predicted sigma points and the corresponding measurement vectors \cite{Julier2000}. Then the conditional mean state vector and the error covariance is computed using the Kalman Filter equations \cite{Julier2000}.

\subsection*{Single propagation Unscented Kalman Filter}
In the SPUKF, only $\bm{X}_0(t)$ is propagated to the next epoch. The other sigma points are not propagated. To calculate the sigma points at the next epoch $t + \delta t$, the following equation is utilized \cite{Biswas2016}
\begin{equation}
\bm{X}_i^-(t+\delta t) = \bm{X}_0^-(t+\delta t) + e^{\bm{\mathcal{J}}\delta t}{\Delta \bm{X}_i}
\label{eq:sigma_approx}
\end{equation}
Here $\bm{X}_0^-(t+\delta t)$ propagated augmented state vector at $t+\delta t$ and
\begin{align}
\bm{\mathcal{J}} & = \left.\frac{\partial \dot{\bm X}_a}{\partial \bm X_a}\right|_{\bm{X}^+_a(t)}\nonumber\\
&= \left[\begin{array}{cc}
\left.\frac{\partial \dot{\bm X}}{\partial \bm X}\right|_{\bm X^+(t)} & \bm{0}_{8\times 8}\\
\bm{0}_{8\times 8} & \bm{0}_{8\times 8}																																	
\end{array}\right]
\label{eq:jacobi}
\end{align}
After calculation of all the sigma points the standard weighted mean and covariance calculation method of the UT \cite{Julier2000} is used to compute the \textit{a priori} mean state vector and the error covariance matrix. The correction stage of the SPUKF is the same as the UKF.

\subsection*{Extrapolated Single propagation Unscented Kalman Filter}
In the ESPUKF, the sigma points are computed using the following equations \cite{Biswas2016}:
\begin{align}
 N_1\bm{ (\Delta X_i)} &= \bm{X}_0^-(t+\delta t) + e^{\bm{\mathcal{J}}\delta t}{\Delta \bm{X}_i}
\label{eq:N1}\\
N_2\bm{(\Delta X_i)} &= \bm{X}_0^-(t+\delta t) + e^{\bm{\mathcal{J}}\delta t}\frac{\Delta \bm{X}_i}{2}
												+ e^{\bm{\mathcal{J}'}\delta t}\frac{\Delta \bm{X}_i}{2}\label{eq:N2}\\
\bm{X}_i^-(t+\delta t) &= 2N_2\bm{(\Delta X_i)} - N_1\bm{ (\Delta X_i)}
\label{eq:richardson}
\end{align}
Here, 
\begin{align}
\bm{\mathcal{J}'} & = \left.\frac{\partial \dot{\bm X}_a}{\partial \bm X_a}\right|_{\bm{X}^+_a(t) + \frac{\Delta \bm{X}_i}{2}}\nonumber
\end{align}
Computation of sigma points using equation \ref{eq:richardson} results in inclusion of the second-order Taylor series terms in the approximation \cite{Biswas2016}. The rest of the calculation procedure in the ESPUKF is the same as for the SPUKF.
\section*{Simulation}
\begin{figure}[h]%
\centering
\includegraphics[width=0.6\columnwidth]{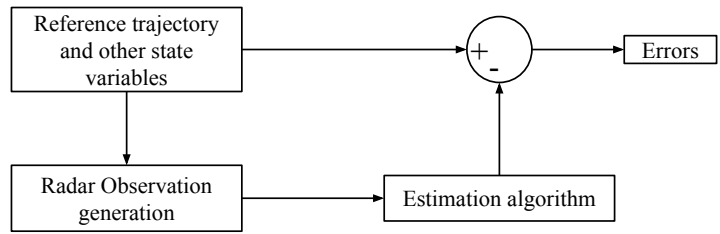}%
\caption{Estimation using radar observations}%
\label{fig:radar_obs}%
\end{figure}

The block diagram for the launch vehicle and re-entry vehicle trajectory estimation experiments using radar observations is shown in Fig. \ref{fig:radar_obs}.

The radar observations corresponding to the reference trajectory are generated using the measurement equations (\ref{eq:crv_range}) and (\ref{eq:crv_elevation}). Measurement noise is also included in the observation simulation. True range and angle measurement noise standard deviation are selected as 50 m and $0.1^{\circ}$. These observations are used in various estimation algorithms to estimate the trajectory of the re-entry vehicle and compared with the reference trajectory for performance analysis. For all the estimation algorithms, the initial state vector and error covariance are:
\begin{equation}
\widehat{\bm X}(0) = \left[\begin{array}{l}
	101~km\\
	5~km\\
	6.050~km/s\\
	-10^{\circ}\\
	0.7
\end{array}\right]
\label{eq:X_ini}
\end{equation}

\begin{equation}
\bm P(0) = diag~\left[\begin{array}{c}
	6\\
	6\\
	.1\\
	.1\\
	.1
\end{array}\right]
\label{eq:P_ini}
\end{equation}
The process noise covariance matrix is selected as
\begin{equation}
\bm Q = 10^{-15}\bm I_{5\times 5}
\label{eq:Q}
\end{equation}

\section*{Simulation Results}
\begin{figure}[h]%
\centering
\begin{subfigure}{0.48\textwidth}
\includegraphics[width=\textwidth]{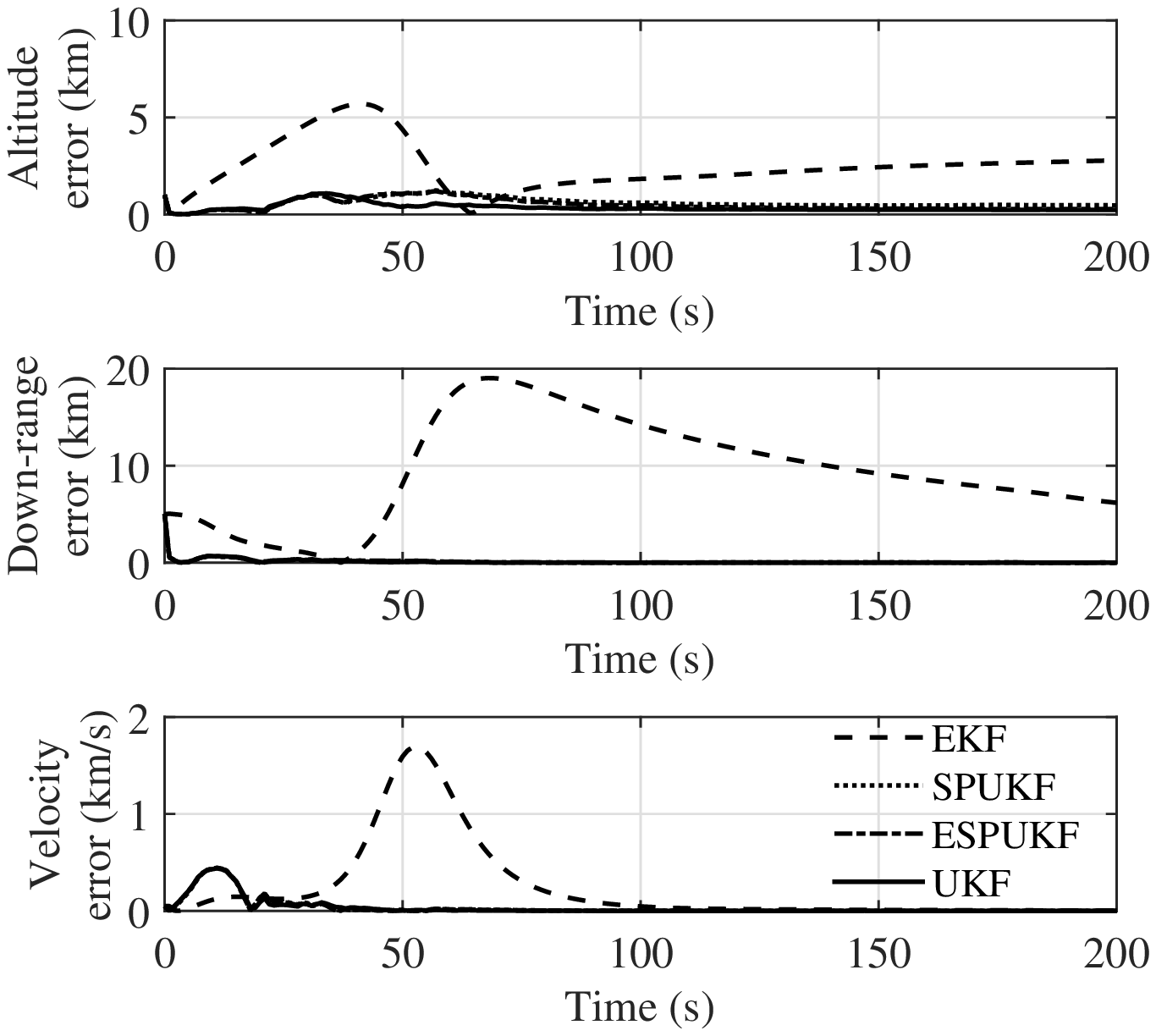}%
\caption{Estimation errors for different KFs}%
\label{fig:crv_KF}%
\end{subfigure}
~
\begin{subfigure}{0.48\textwidth}%
\centering
\includegraphics[width=\textwidth]{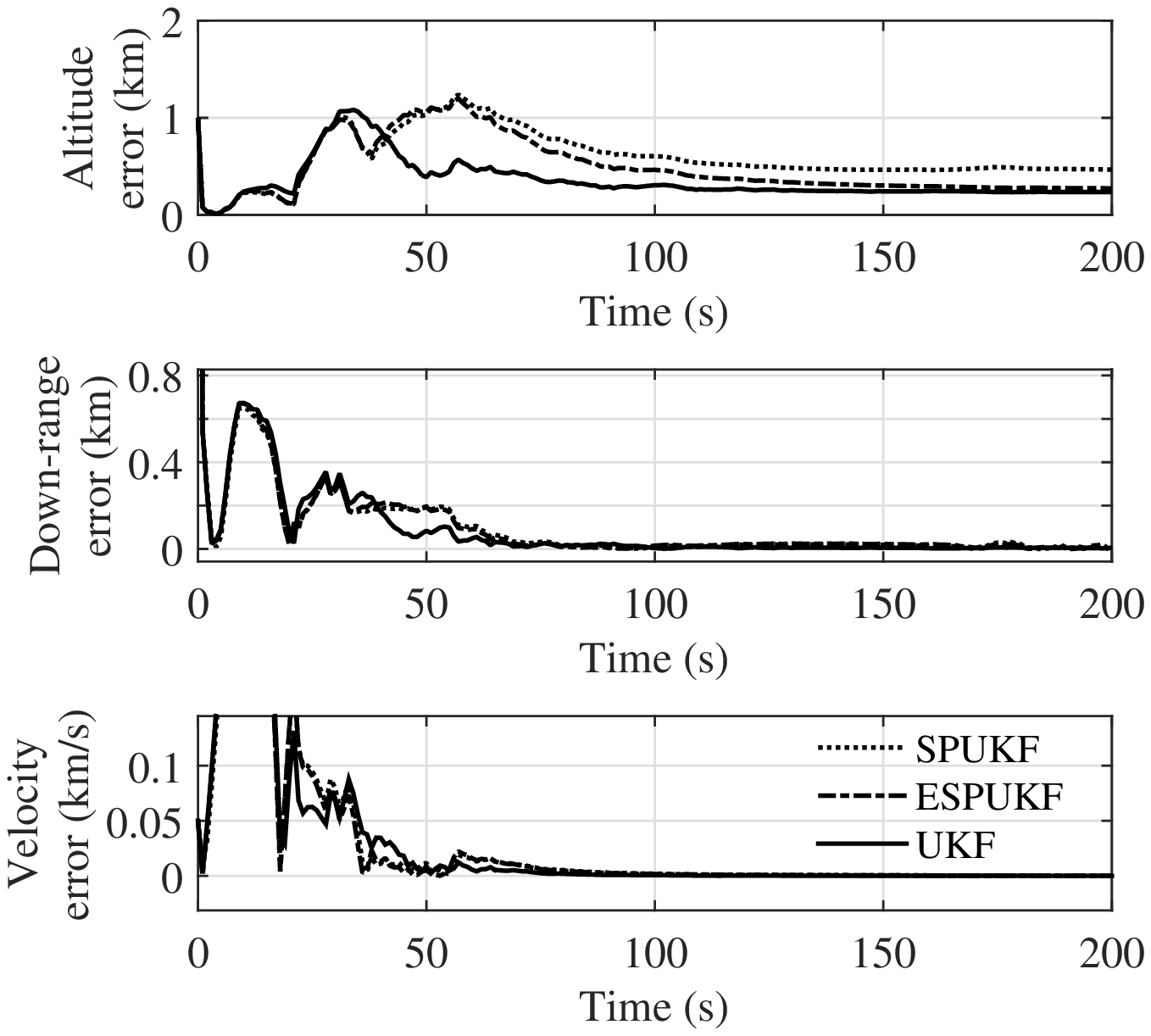}%
\caption{Estimation errors for UKFs}%
\label{fig:crv_UKF}%
\end{subfigure}
\caption{Estimation errors for different algorithms}%
\end{figure}

\begin{figure}%
\centering
\includegraphics[width=0.7\columnwidth]{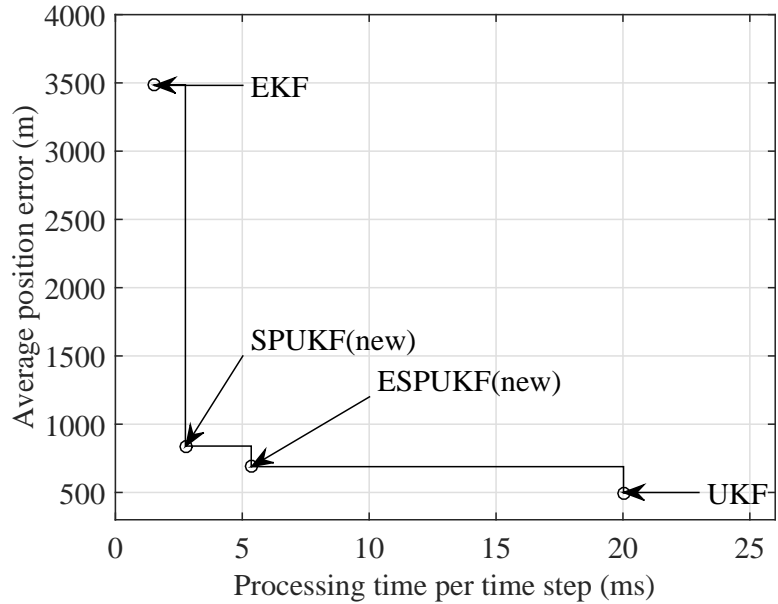}%
\caption{Processing time vs. estimation error}%
\label{fig:RV_KF_PT}%
\end{figure}

The estimation performance of the EKF, UKF, SPUKF and ESPUKF are examined in the re-entry vehicle trajectory estimation scenario. Using the simulation process described in the previous section, the re-entry trajectory is estimated using all the four Kalman Filters separately. The altitude, down-range and velocity errors are shown in Fig. \ref{fig:crv_KF}.
It can be observed that the estimation error of the EKF is much higher than the UKF and the new variants of the UKF. Estimation errors for the UKF, SPUKF and ESPUKF are shown separately in figure \ref{fig:crv_UKF}. The UKF provides the most accurate state estimate of all the Kalman Filters under consideration. The performance of the SPUKF is significantly better than the EKF. The performance of the ESPUKF almost matches the UKF. The processing time required in every time step is also recorded for all the estimation algorithms. The time average estimation error vs. the processing time is shown in Fig. \ref{fig:RV_KF_PT}. The processing time of the EKF is the lowest. However the average estimation error of the EKF is significantly more than the SPUKF, ESPUKF and the UKF. The UKF provides the lowest average estimation error and the processing time is significantly higher than the other Kalman Filters. 
\section*{Conclusion}
In this paper the application of two new variants of Unscented Filters is proposed for the re-entry vehicle trajectory estimation. The results demonstrate that the UKF provides better accuracy than the EKF but requires more processing time. The SPUKF accuracy is better than the EKF and the processing time is significantly less than for the UKF. The processing time reduces by 86.2\% for the SPUKF compared to the UKF. However, the accuracy of the SPUKF is less than the UKF. The ESPUKF provides estimation accuracy comparable to the UKF and the processing time reduces by 73.3\% compared to the UKF.


\bibliographystyle{myieeetr}
\bibliography{biblog}
\end{document}